\documentclass[journal=jctcce,amsmath,amssymb,manuscript=article]{achemso}

\usepackage[utf8]{inputenc}
\usepackage{graphicx}
\usepackage{mathtools}
\usepackage{xcolor}
\usepackage{amsmath}
\usepackage{amssymb}
\usepackage{xfrac}
\usepackage{soul}

\title{Hybrid Gauge Approach for Accurate Real-Time TDDFT Simulations with Numerical Atomic Orbitals}
\author{Haotian Zhao}
\affiliation{CAS Key Laboratory of Quantum Information, University of Science and Technology of China, Hefei 230026, People's Republic of China}

\author{Lixin He}
\email{helx@ustc.edu.cn}
\affiliation{CAS Key Laboratory of Quantum Information, University of Science and Technology of China, Hefei 230026, People's Republic of China}
\affiliation{Institute of Artificial Intelligence, Hefei Comprehensive National Science Center, Hefei, 230088, People's Republic of China}
\affiliation{Hefei National Laboratory, University of Science and Technology of China, Hefei 230088, China}

\date{\today}

\begin{document}

\begin{abstract}
Ultrafast real-time dynamics are critical for understanding a broad range of physical processes. Real-time time-dependent density functional theory (rt-TDDFT) has emerged as a powerful computational tool for simulating these dynamics, offering insight into ultrafast processes and light–matter interactions. In periodic systems, the velocity gauge is essential because it preserves the system’s periodicity under an external electric field. Numerical atomic orbitals (NAOs) are widely employed in rt-TDDFT codes due to their efficiency and localized nature.
However, directly applying the velocity gauge within the NAO basis set neglects the position-dependent phase variations within atomic orbitals induced by the vector potential, leading to significant computational errors—particularly in current calculations. To resolve this issue, we develop a hybrid gauge that incorporates both the electric field and the vector potential, preserving the essential phase information in atomic orbitals and thereby eliminating these errors. Our benchmark results demonstrate that the hybrid gauge fully resolves the issues encountered with the velocity gauge in NAO-based calculations, providing accurate and reliable results. This algorithm offers a robust framework for future studies on ultrafast dynamics in periodic systems using NAO bases.
\end{abstract}
\maketitle

\section{Introduction}

Ultrafast real-time dynamics are crucial for understanding a wide range of complex physical processes. \cite{Zewail2000,Strickland1985,Krausz2009,Corkum2007,Schultze2010,Huillier1993,
Goulielmakis2008,zurch2014,Chini2014}
In fields such as materials science, chemical reaction dynamics, and condensed matter physics, the ability to track electron motion on attosecond (as) to femtosecond (fs) timescales is essential for understanding phenomena such as photoconversion, ultrafast optical responses, and energy transfer.\cite{Krausz2009,Corkum2007} Exploring these ultrafast processes has become a key driving force in advancing these research areas.

Real-time time-dependent density functional theory (rt-TDDFT) \cite{Runge1984,Yabana1996,Bertsch2000,Marques2004,Yabana1999,Castro2004td,Otobe2008,Provorse2016,Sato2014}
 directly simulates a system's time evolution under external fields or time-dependent interactions, capturing instantaneous electronic responses. It has become an indispensable theoretical tool for investigating ultrafast dynamics, nonlinear phenomena, light-matter interactions, nonequilibrium states, and excited-state dynamics, etc.

When studying the interaction between isolated systems and light fields, the length gauge is typically employed, where the interaction is expressed through $\boldsymbol{E }\cdot \boldsymbol{r}$. However, in periodic systems, the $\boldsymbol{E }\cdot \boldsymbol{r}$ form breaks the periodic symmetry of the system, making the length gauge unsuitable. In such cases, the velocity gauge,\cite{Bertsch2000,Yabana2012} which introduces the vector potential $\boldsymbol{A}$ instead of the scalar potential, becomes a more appropriate method. This approach preserves the system's periodicity and is particularly well-suited for simulating ultrafast electron dynamics in periodic systems.

Numerical atomic orbitals (NAOs) offer several advantages as basis sets in electronic structure calculations.\cite{NAO-review} Compared to other types of basis sets, NAOs generally require fewer basis functions, significantly reducing computational costs.
The rt-TDDFT based on NAO and velocity gauge has been implemented in several codes.\cite{PEMMARAJU201830,TDAP,Aims-tddft,Hefx2024}
However, we show that directly applying the velocity gauge within the NAO basis set neglects the position-dependent phase variations within atomic orbitals caused by the vector potential, leading to computational errors. This issue is particularly problematic when calculating current, where these errors can become significant and sometimes even result in qualitatively incorrect results.

To address the errors arising from the use of the velocity gauge within the NAO basis set,
we develop a hybrid gauge algorithm. This method employs both the electric field and the vector potential,
combining the strengths of the length and velocity gauges by retaining crucial position-dependent
phase information within atomic orbitals. By resolving the computational errors of the velocity gauge,
particularly in current calculations, the hybrid gauge significantly improves simulation accuracy.
Furthermore, the hybrid gauge is more computationally efficient than the velocity gauge,
as it does not require time-consuming updates of nonlocal pseudopotentials.
This efficiency, together with the minimal modifications needed to integrate it into
the existing length-gauge framework, makes the hybrid gauge well-suited for practical implementation
in established computational codes. Benchmark results demonstrate its robustness and versatility,
offering a powerful tool for investigating ultrafast dynamics and other advanced phenomena.

\section{Methods}

In rt-TDDFT, the key is to solve the time-dependent Kohn-Sham (TDKS) equation\cite{Lian2018,You2021}:
\begin{equation}
i\frac{\partial}{\partial t}{\psi_{n}(\boldsymbol{r},t)} = H \psi_{n}(\boldsymbol{r},t),
\end{equation}
where $H$ and $\psi_{n}$ are the Kohn–Sham Hamiltonian and orbitals, respectively.

For finite systems subjected to an external electric field, the Hamiltonian in the length gauge is expressed as:
\begin{equation}
H = H_0 + \boldsymbol{E}(t) \cdot \boldsymbol{r},
\end{equation}
where $H_0$ represents the Hamiltonian of the system without an external field, given by:
\begin{equation}
H_0 = -\frac{1}{2}\nabla^2 + V_{\rm Hxc}(\boldsymbol{r}) + V^{\rm ps}_{\rm L}(\boldsymbol{r}) + V^{\rm ps}_{\rm NL}.
\end{equation}
Here, $V_{\rm Hxc}(\boldsymbol{r})$ is the Hartree-exchange-correlation potential, while $V^{\rm ps}_{\rm L}(\boldsymbol{r})$ and $V^{\rm ps}_{\rm NL}$ denote the local and non-local components of the pseudopotential, respectively. The non-local pseudopotential, $V^{\rm ps}_{\rm NL}$, is commonly expressed in the Kleinman-Bylander (KB) form \cite{Kleinman1982}.
The term $\boldsymbol{E}(t) \cdot \boldsymbol{r}$ represents the external potential induced by a time-dependent electric field, which is often assumed to be spatially uniform in simulations.

However, because the operator $\boldsymbol{E} \cdot \boldsymbol{r}$ does not preserve translational symmetry, it cannot be used in periodic systems. In such cases, the velocity-gauge TDKS equation must be employed. To apply an external electric field under periodic boundary conditions, a vector potential is introduced \cite{PEMMARAJU201830},
\begin{equation}
\boldsymbol{E}(t) = -\frac{\partial \boldsymbol{A}(t)}{\partial t}.
\end{equation}
The vector potential $\boldsymbol{A}$ is assumed to be spatially uniform within the system.
The TDKS equation in the velocity gauge is given by
\begin{equation}
i\frac{\partial}{\partial t}\widetilde{\psi}_{n}(\boldsymbol{r},t) = \widetilde{H}\widetilde{\psi}_{n}(\boldsymbol{r},t),
\label{eq:tdks_v}
\end{equation}
where the Hamiltonian is
\begin{equation}
\widetilde{H} = \frac{1}{2}\left[-i \nabla + \boldsymbol{A}(t)\right]^2 + V_{Hxc} + V^{\rm ps}_{\rm L} + \widetilde{V}^{\rm ps}_{\rm NL}.
\label{eq:ham_vel}
\end{equation}
We use `` $\widetilde{}$ '' to denote quantities in the velocity gauge.
The wave functions in the velocity gauge are related to those in the length gauge via a gauge transformation: \cite{Bertsch2000,Yabana2012}
\begin{equation}
\widetilde{\psi}_n(\boldsymbol{r}, t) = e^{-i \boldsymbol{A}(t) \cdot \boldsymbol{r}} \psi_n(\boldsymbol{r}, t).
\label{eq:psi_len_vel}
\end{equation}
The non-local pseudopotential term in the velocity gauge is given by \cite{PEMMARAJU201830}
\begin{equation}
\widetilde{V}^{\rm ps}_{\rm NL} \widetilde{\psi}_{n}(\mathbf{r}, t) = \int d\mathbf{r}^{\prime} \, e^{-i \boldsymbol{A}(t) \cdot \mathbf{r}} V_{\mathrm{\rm NL}}(\mathbf{r}, \mathbf{r}^{\prime}) e^{i \boldsymbol{A}(t) \cdot \mathbf{r}^{\prime}} \widetilde{\psi}_{n}\left(\mathbf{r}^{\prime}, t\right).
\end{equation}
This part is computationally expensive, as it involves integration over grid points in real space.


In the NAO basis sets, the time-dependent wave functions in the length gauge can be written as a linear combination of atomic orbitals (LCAO)\cite{Soler/etal:2002,ABACUS2,NAO-review},
\begin{equation}
\psi_{n }(\boldsymbol{r}, t) = \sum_\mu c_{n \mu }(t) \phi_{\mu }(\boldsymbol{r})\, .
\label{eq:psi_len}
\end{equation}
The TDKS equation can then be expressed in matrix form as \cite{Lian2018,You2021}:
\begin{equation}
i \sum_\nu S_{\mu \nu}(t) \frac{\partial c_{n \nu }(t)}{\partial t} = \sum_\nu H_{\mu \nu }(t) c_{n \nu }(t)\, ,
\label{eq:tdks-length}
\end{equation}
where $S_{\mu \nu} = \langle \phi_\mu |\phi_\nu \rangle$ is the overlap matrix element between the NAOs.

Similarly, the wave functions in velocity gauge on the NAO basis are written as,
\begin{equation}
\tilde{\psi}_{n }(\boldsymbol{r}, t) = \sum_\mu \tilde{c}_{n \mu }(t) \phi_{\mu }(\boldsymbol{r})\, .
\label{eq:psi_vel}
\end{equation}
However, comparing Eqs.~(\ref{eq:psi_len_vel}), (\ref{eq:psi_len}), and (\ref{eq:psi_vel}), it is evident that the NAO wave functions in the velocity gauge neglect the position-dependent phases, as $\tilde{c}_{n \mu }(t)$ depends only on time.
We shall demonstrate that neglecting these phases may result in significant errors in the calculations, especially in the case of current calculations.

To address this issue, we propose a new scheme, which is  named hybrid gauge,
for performing rt-TDDFT calculations using NAO basis sets.
The derivation of the hybrid gauge involves substantial mathematical detail. To maintain clarity and keep the main text focused, we present only the key steps and essential equations here. The full derivation is provided in the Supporting Information (SI) for further reference.

In the proposed algorithm, we use a bar (instead of a tilde) over the wave functions and related terms to distinguish them from those in traditional velocity-gauge algorithms.
We write the Bloch wave functions at a given momentum $\boldsymbol{k}$ as,
\begin{equation}
\bar{\psi}_{n\boldsymbol{k}}(\boldsymbol{r}, t)
= \sum_\mu \bar{c}_{n \mu \boldsymbol{k}}(t)\,\bar{\phi}_{\mu \boldsymbol{k}}(\boldsymbol{r}, t),
\label{eq:psi_hyb}
\end{equation}
where the basis functions in the hybrid gauge are given by
\begin{equation}
\bar{\phi}_{\mu \boldsymbol{k}}(\boldsymbol{r}, t)
= \frac{1}{\sqrt{N}} \sum_{\boldsymbol{R}_i}
  e^{i \boldsymbol{k} \cdot \boldsymbol{R}_i}\,
  \bar{\phi}_\mu \!\bigl(\boldsymbol{r} - \boldsymbol{\tau}_\mu - \boldsymbol{R}_i, t\bigr).
\end{equation}
Here, $\bar{\phi}_{\mu}\!\bigl(\boldsymbol{r} - \boldsymbol{\tau}_\mu - \boldsymbol{R}_i, t \bigr)
= e^{-i \boldsymbol{A}(t) \cdot \Delta \boldsymbol{r}_{\mu i}}
  \,\phi_{\mu}\!\bigl(\boldsymbol{r} - \boldsymbol{\tau}_\mu - \boldsymbol{R}_i\bigr)$,
with $\Delta \boldsymbol{r}_{\mu i} = \boldsymbol{r} - \boldsymbol{\tau}_{\mu} - \boldsymbol{R}_i$, and $\boldsymbol{\tau}_{\mu}$ being the center of the $\mu$-th orbital.
This modification maintains the phase information within the NAOs.

The time evolution of the Bloch wave function is then given by:
\begin{equation}
i \sum_\nu \bar{S}_{\mu \nu}(\boldsymbol{k}, t) \frac{\partial \bar{c}_{n \nu \boldsymbol{k}}(t)}{\partial t} = \sum_\nu \bar{H}_{\mu \nu}(\boldsymbol{k}, t) \bar{c}_{n \nu \boldsymbol{k}}(t),
\label{eq:tdks_v_p}
\end{equation}
where
$$
\bar{S}_{\mu \nu}(\boldsymbol{k}, t) = \langle \bar{\phi}_{\mu \boldsymbol{k}}(t) | \bar{\phi}_{\nu \boldsymbol{k}}(t) \rangle,
$$
is the overlap matrix in the hybrid gauge. The Hamiltonian is given by,
\begin{equation}
\bar{H}_{\mu \nu}(\boldsymbol{k}, t) = \sum_{\boldsymbol{R}_i} e^{i \boldsymbol{k} \cdot \boldsymbol{R}_i} \, \bar{H}_{\mu \nu}(\boldsymbol{R}_i),
\label{eq:Hk_hyb}
\end{equation}
where
\begin{equation}
\bar{H}_{\mu \nu}(\boldsymbol{R}_i) = e^{-i \boldsymbol{A}(t) \cdot \boldsymbol{\tau}_{\mu 0, \nu i}}
\langle \phi_{\mu 0} \,|\, H_0 + \boldsymbol{E}(t) \cdot \Delta \boldsymbol{r}_{\nu i} \,|\, \phi_{\nu i} \rangle.
\end{equation}
Here, $ \boldsymbol{\tau}_{\mu 0, \nu i} = \boldsymbol{\tau}_{\mu} - \boldsymbol{\tau}_{\nu}  - \boldsymbol{R}_i $
is the displacement vector between the two atoms.
The overlap matrix in the hybrid gauge can be calculated in a similar way.

Compared to the length gauge, the Hamiltonian in the hybrid gauge introduces two main changes. First, it includes an additional phase factor that depends on the relative positions between atoms.
Second, the term $\boldsymbol{E}(t) \cdot \boldsymbol{r}$ in the length gauge is replaced by $\boldsymbol{E}(t) \cdot \Delta \boldsymbol{r}_{\nu i}$.
It is noteworthy that the integral involving $\boldsymbol{E}(t) \cdot   \boldsymbol{r}$ is non-periodic and unbounded in periodic systems. In contrast, $\boldsymbol{E}(t) \cdot \Delta \boldsymbol{r}_{\nu i}$ avoids these issues, as it describes the local displacement of electrons relative to an atomic center, leading to a bounded integral.

The time-dependent current is one of the important quantities that can be calculated from the TDKS equations. It can further be used to compute optical spectra and dielectric functions.
The time-dependent current density in the length gauge is given by:
\begin{equation}
\boldsymbol{j}(t) = \frac{-1}{2\Omega N_k} \sum_{n\boldsymbol{k}} f_{n\boldsymbol{k}}
\left[\left\langle \psi_{n \boldsymbol{k}} \middle| \boldsymbol{\pi} \middle| \psi_{n \boldsymbol{k}} \right\rangle + \text{c.c.}\right],
\end{equation}
where $\Omega$ is the volume of the unit cell, $N_k$ is the number of $k$ points, and
\begin{equation}
\boldsymbol{\pi} = \frac{1}{i}\left[\boldsymbol{r}, H\right] = -i \nabla + i \left[V^{\rm ps}_{\rm NL}, \boldsymbol{r}\right].
\end{equation}
is the generalized momentum.
For the velocity gauge, $\widetilde{\boldsymbol{\pi}} =-i \nabla + \boldsymbol{A}(t)+ i \left[\widetilde{V}^{\rm ps}_{\rm NL}, \boldsymbol{r}\right]$  \cite{PEMMARAJU201830}.

After some derivation, we have the expression of the current density in the hybrid gauge:
\begin{eqnarray}
\boldsymbol{j}(t) &=& \frac{-1}{\Omega N_k}  \sum_{\boldsymbol{k}} \operatorname{Re}\left[ \sum_{\mu \nu}
\bar{\rho}_{\nu \mu}(\boldsymbol{k})
\right. \nonumber \\
&& \left. \sum_{\boldsymbol{R}} e^{i \boldsymbol{k} \cdot \boldsymbol{R}} e^{-i \boldsymbol{A}(t) \cdot \left(\Delta \tau_{\mu \nu} - \boldsymbol{R}\right)}
\left\langle \phi_{\mu 0} \middle|  \boldsymbol{\pi}
\middle| \phi_{\nu \boldsymbol{R}} \right\rangle \right].
\end{eqnarray}
where,
\begin{equation}
\bar{\rho}_{\mu\nu} (\boldsymbol{k}) = \sum_{n} f_{n\boldsymbol{k}} \bar{c}_{n\mu}(\boldsymbol{k}) \bar{c}_{n\nu}^*(\boldsymbol{k})
\end{equation}
 is the density matrix in the hybrid gauge.

The hybrid gauge offers several advantages over the velocity gauge. First, it preserves the position-dependent phase information within NAOs,  which is expected to enhance accuracy. Additionally, it requires only minimal modifications to the length gauge, making it straightforward to implement. Notably, whereas the velocity gauge requires time-consuming grid-point integration to compute matrix elements of non-local potentials at each time step, the hybrid gauge does not. Instead, in both the length and hybrid gauges, these calculations can be handled using two-center integrals, which do not need to be updated as long as the ions remain fixed, thereby boosting computational efficiency.

\section{Results and Discussion}

\begin{figure}[tbp]
	\centering
	\includegraphics[width=0.75\linewidth]{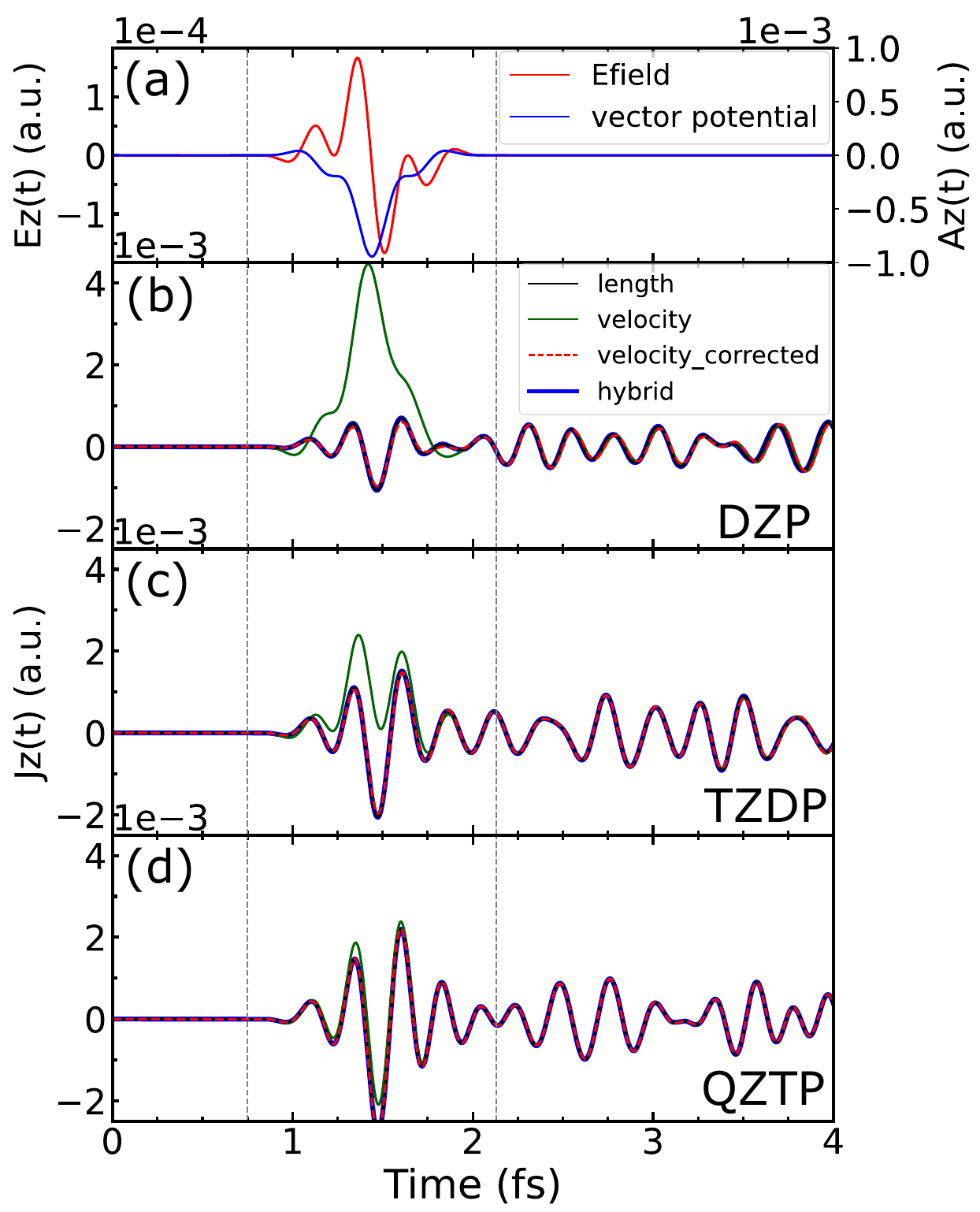}
	\caption{Laser-induced time-dependent currents for H$_2$CO$_3$: (a) Applied electric field and vector potential; comparison of time-dependent currents from the length, hybrid, velocity, and corrected velocity gauges using (b) DZP, (c) TZDP, and (d) QZTP basis sets.}
\label{H2CO3 current}
\end{figure}

We have implemented the length gauge, velocity gauge, and hybrid gauge rt-TDDFT in the ABACUS code on NAO basis. \cite{ABACUS2,Lin_orbital} In the following sections, we compare the results obtained from these three gauges. First, we analyze the optical spectra of molecular systems. Next, we compare the results for periodic systems, using bulk silicon as an example.

\subsection{Optical response in Carbonic acid Molecule}

\begin{figure}[htbp]
	\centering
	\includegraphics[width=0.75\linewidth]{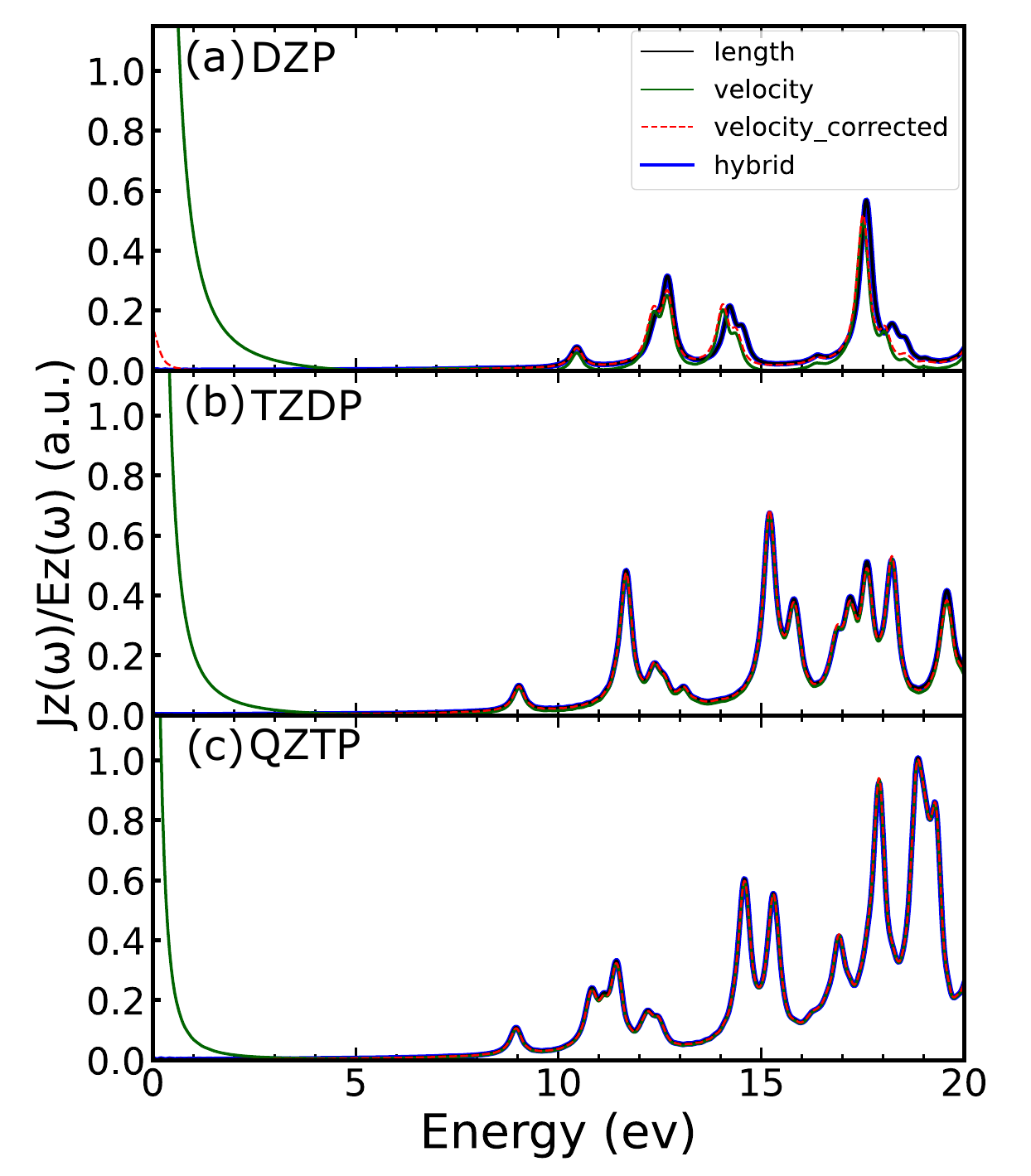}
	\caption{Comparison of  absorption spectra of H$_2$CO$_3$ from the length, hybrid, velocity, and corrected velocity gauges using (a) DZP, (b) TZDP, and (c) QZTP basis sets.}
	\label{H2CO3 spectrum}
\end{figure}

We first compare the optical spectra of molecules. We use the Perdew-Burke-Ernzerhof (PBE) exchange-correlation functional~\cite{GGA} and SG15 norm-conserving pseudopotentials,~\cite{SG15} and we employ NAO basis sets to expand the Kohn-Sham wave functions. We apply a Gaussian wave packet modulating two monochromatic light waves (see SI), as shown in Fig.~\ref{H2CO3 current}(a). The packet is centered at $t_0$= 1.44 fs, and the standard deviation is $\sigma$= 0.2 fs.
We carry out the time evolution with a step size of 2.4~as over 8000 steps (totaling 19.2~fs), while keeping the ions fixed. We obtain the vector potentials for the velocity gauge and hybrid gauge by numerically integrating the electric field over time.

We benchmark a series of small molecular systems, using the H$_2$CO$_3$ molecule as an illustrative example. The remaining results are provided in the SI. To examine convergence with respect to the basis set size, we compare
the results of three NAO basis sets: double-$\zeta$ plus polarization functions (DZP), triple-$\zeta$ plus double-polarization functions (TZDP), and quadruple-$\zeta$ plus triple-polarization functions (QZTP).~\cite{Lin_orbital} The calculated time-dependent currents and optical spectra of H$_2$CO$_3$ are shown in Fig.~\ref{H2CO3 current} and Fig.~\ref{H2CO3 spectrum}, respectively.

Figure~\ref{H2CO3 current}(b) depicts the results for the DZP basis set. The time-dependent currents from the length gauge and the hybrid gauge are essentially identical, differing only by negligible numerical errors. Consequently, the optical spectra calculated using both gauges are also indistinguishable, as shown in Fig.~\ref{H2CO3 spectrum}(a).

However, the current calculated using the velocity gauge deviates significantly from the other two gauges during the laser field duration, which leads to an unphysical low-frequency divergence in the absorption spectrum, as shown in Fig.~\ref{H2CO3 spectrum}(a). Additionally, some discrepancies are observed in the optical spectra. We find that the deviation in the current is proportional to the magnitude of the vector field. We therefore propose a numerical correction method (see SI for details). After applying the correction, the current obtained with the velocity gauge becomes nearly identical to that of the other two gauges. Consequently, the unphysical low-frequency features are greatly reduced, and the overall spectra agree very well with those from the other two gauges, although small discrepancies remain. One reason the results of the correction are excellent is that the electric field is only applied for a very short period, and for most of the time, the system evolves under zero electric field.

When the TZDP basis set is used, the deviation of the velocity-gauge current from those of the other two gauges is significantly reduced, though it still remains substantial, as shown in Fig.~\ref{H2CO3 current}(b). This suggests that a larger basis set offers greater flexibility to accommodate phase changes within the molecules. In the spectra, the unphysical low-frequency divergence persists, but aside from this artifact, the overall spectra are identical to those obtained using the length gauge and the hybrid gauge. After the correction is applied, the low-frequency divergence disappears completely.

When the basis set is further increased to QZTP, the results improve even more. The deviation in the current becomes smaller, even before applying the correction. However, the low-frequency divergence in the spectra still persists, yet it disappears after the correction is applied to the current.

The above results demonstrate that the length gauge and the hybrid gauge yield identical time-dependent currents and optical spectra. In contrast, the velocity gauge shows significant deviations in the current during the laser field, leading to an unphysical low-frequency divergence in the absorption spectra. Increasing the basis set size reduces these deviations. The numerical corrections significantly improve the results of the velocity gauge. All other molecules we tested exhibit the same behavior.

\begin{figure}[tbp]
	\centering
	\includegraphics[width=0.70\linewidth]{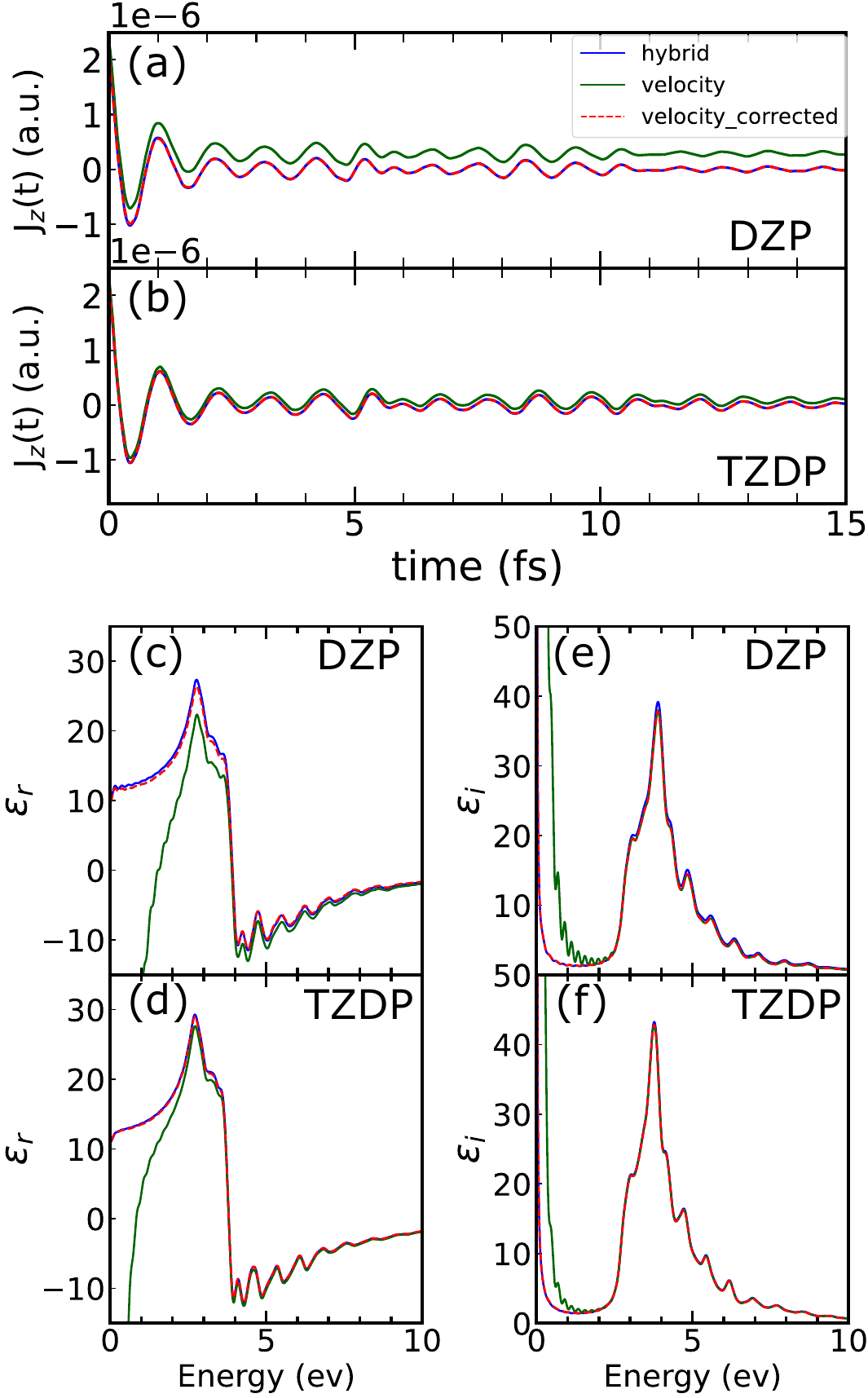}
	\caption{
(a) and (b) show the currents under a $\delta$-function electric field obtained using the DZP and TZDP basis sets for bulk Si; (c) and (d) display the real part of the dielectric function of Si using the DZP and TZDP basis sets; (e) and (f) show the imaginary part of the dielectric function of Si using the DZP and TZDP basis sets. The blue solid line, green solid line, and red dashed line correspond to the hybrid, velocity, and corrected velocity gauges, respectively.
}
	\label{linear_Si}
\end{figure}

\subsection{Field–matter interactions in bulk Silicon}
In the previous section, we demonstrate the effectiveness of the hybrid gauge for molecular spectra.
For periodic systems, however, because the electric potential field breaks the periodicity, the velocity gauge
is commonly considered the only valid choice.
Here, we perform a series of comparative tests on the velocity and hybrid gauges in periodic systems
to show that the hybrid gauge can provide more accurate and reliable results than the velocity gauge.

We first examine the optical response functions of bulk silicon. Our simulation system consists of an 8-atom cubic conventional cell of bulk silicon. A $\Gamma$-centered 16$\times$ 16 $\times$ 16 $k$-point mesh is used to ensure
the convergence of the results. We apply a $\delta$-function electric field at the first step, which generates a step-like vector potential with a magnitude of $-8\times 10^{-5}$~a.u. We use the PBE exchange-correlation functional , along with DZP (2s2p1d) and TZDP (3s3p2d) basis sets for the Si element.~\cite{Lin_orbital}  The cutoff radius is 8~a.u. We run the simulation for 10,000 steps over a total of 19.3~fs, using a time step of 1.93~as.

The time-dependent currents for the hybrid gauge and velocity gauge are shown in Fig.~\ref{linear_Si}(a) and (b) for the DZP and TZDP basis sets, respectively. Similar to the molecular system, the current calculated using the velocity gauge shows a significant discrepancy compared to the hybrid gauge before applying the correction. However, after the correction, the two results align well. For the TZDP basis set, the velocity gauge results are closer to the hybrid gauge results compared to the DZP basis set, both before and after the correction.

Figure~\ref{linear_Si}(c) and (e) show the real and imaginary parts of the dielectric functions for the DZP basis set, while Figure~\ref{linear_Si}(d) and (f) present the corresponding results for the TZDP basis set. Notably, for the hybrid gauge, even the small DZP basis set produces dielectric functions in excellent agreement with those of the TZDP basis set, which are also in good agreement with the results reported in Ref.~\cite{PEMMARAJU201830} using the ARTED code.
In contrast, for both basis sets, the dielectric functions calculated using the velocity gauge deviate significantly from those obtained with the hybrid gauge, and exhibit unphysical behavior at low frequencies before the correction. After applying the correction, the dielectric functions show significant improvement and align well with those obtained from the hybrid gauge. Notably, Ref.~\cite{PEMMARAJU201830} employs a significantly larger basis set, with each Si atom represented by a total of 27 atomic orbitals within the NAO scheme, compared to 13 for DZP and 22 for TZDP bases in our calculations.

\begin{figure}[tbp]
	\centering
	\includegraphics[width=1.1\linewidth]{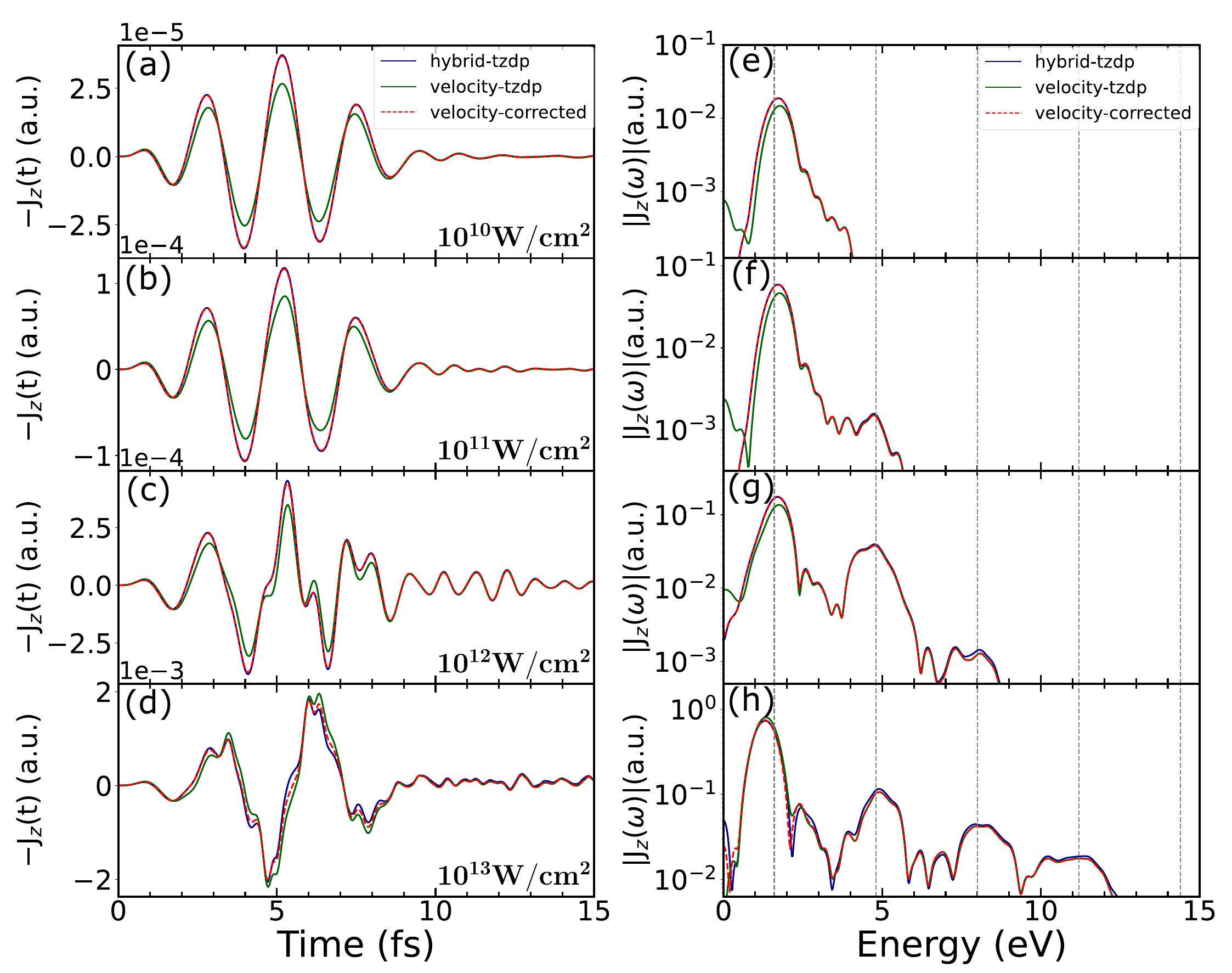}
	\caption{Laser-induced nonlinear response with bulk Si. (a), (b), (c), (d) represent the currents under different field intensities ranging from $10^{10}$ W/cm$^2$ to $10^{13}$ W/cm$^2$; (e), (f), (g), (h) show the corresponding results after Fourier transformation into the frequency domain. In (e), (f), (g), and (h), the 1st, 3rd, 5th, 7th, and 9th harmonics are marked with gray vertical lines.
}
	\label{nonlinear}
\end{figure}

\begin{figure}[tbp]
	\centering
	\includegraphics[width=0.75\linewidth]{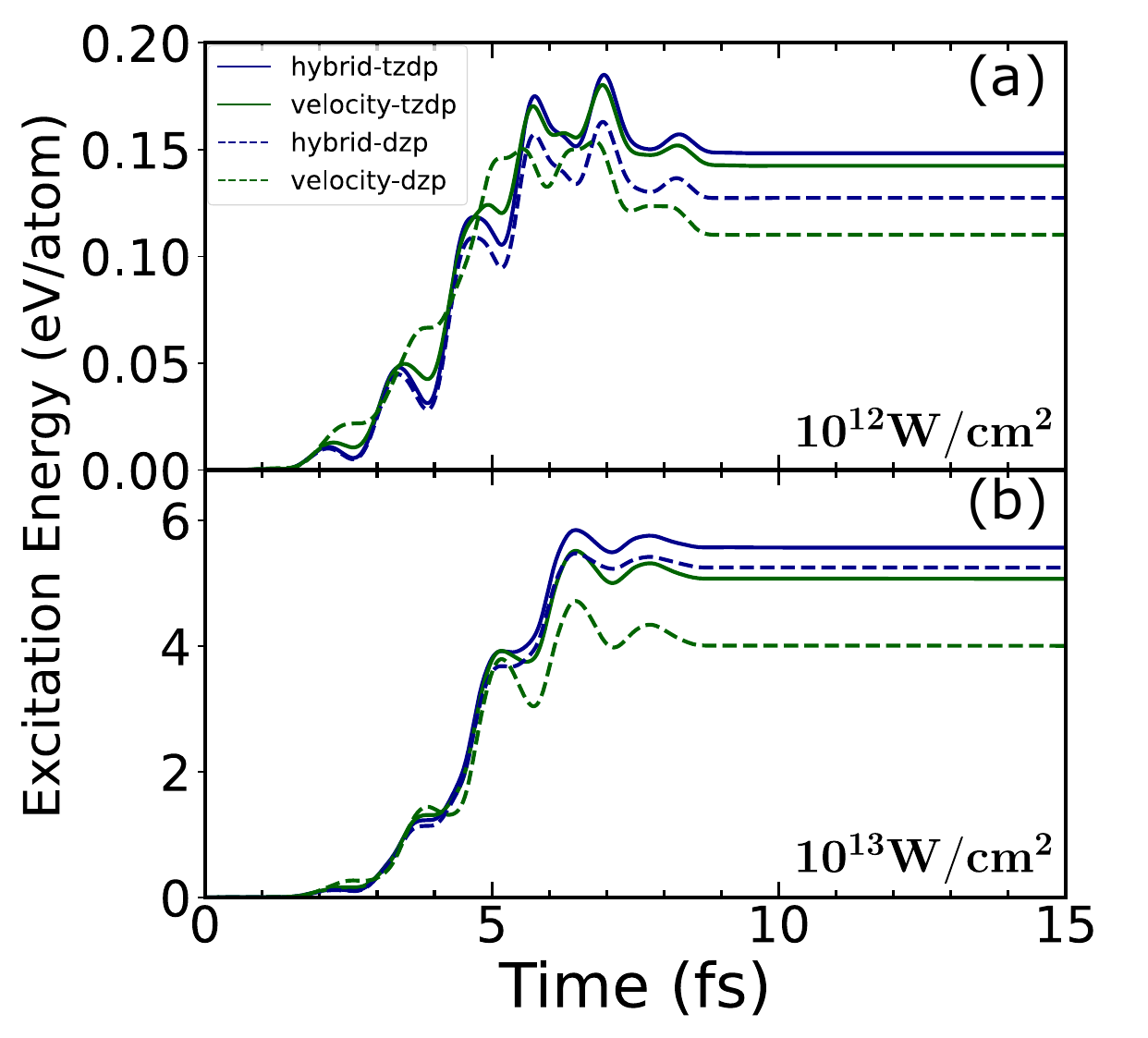}
	\caption{The variation of excitation energy over time under Laser fields in bulk Si with intensities of (a) $10^{12}$ W/cm$^2$ and (b) $10^{13}$ W/cm$^2$.
}
	\label{fig:Excitation}
\end{figure}

To further demonstrate the effectiveness of the hybrid gauge in strong fields, we investigate nonlinear effects under strong-field conditions. The laser field employed is a sin$^{2}$ envelope centered at 5~fs with a carrier frequency of 1.6~eV, to facilitate comparison with the results from Ref.~\cite{PEMMARAJU201830}. The excitation power ranges from 10$^{10}$~W/cm$^2$ to 10$^{13}$~W/cm$^2$. The calculated currents in the time and frequency domains are shown in Fig.~\ref{nonlinear}. In the main text, we present only the results obtained using the TZDP basis set.

The overall trends of the currents calculated with both gauges are generally consistent, with both showing increasingly pronounced nonlinear effects as the electric field strength increases. When the electric field strength reaches $10^{13}$~W/cm$^2$, the seventh harmonic is clearly observed, which agrees well with Ref.~\cite{PEMMARAJU201830}. However, as before, some discrepancies are observed between the two gauges prior to correction.
Interestingly, the results of the hybrid gauge closely match those of the ARTED code~\cite{ARTED} reported in Ref.~\cite{PEMMARAJU201830}, whereas the results of the velocity gauge are similar to the results obtained with the SIESTA code in Ref.~\cite{PEMMARAJU201830}, despite the much larger basis set used in the SIESTA calculations.
After applying the correction to the velocity gauge, the results from the two gauges align closely.
However, the numerical correction cannot fully eliminate the errors. As shown in Fig.~\ref{nonlinear}(d) and (h), differences between the currents of the two gauges persist even after correction. These differences become more pronounced with smaller basis sets and stronger electric fields, leading to significant errors in the frequency-domain results, as detailed in the
SI.

Figure \ref{fig:Excitation} presents the excitation energies under strong laser fields \cite{PEMMARAJU201830}. Our results align with the trends reported in the literature, showing oscillations that approximately follow the half-cycles of the laser field and stabilizing after the pulse ends, as depicted in Fig.~\ref{fig:Excitation}. At a field strength of 10$^{12}$W/cm$^2$, the final excitation energy calculated with the hybrid gauge using the TZDP basis set is approximately 0.02 eV/atom lower than that of the ARTED code reported in Ref.~\cite{PEMMARAJU201830}, while the result with the DZP basis set is even lower. This difference is primarily due to the fewer basis functions employed in our calculations compared to Ref.~\cite{PEMMARAJU201830}. At 10$^{13}$W/cm$^2$,
our results also close match those in Ref.~\cite{PEMMARAJU201830}.
In comparison, the excitation energies computed using the velocity gauge are generally lower, with accuracy significantly deteriorating for smaller basis sets and at higher field strengths. Unlike currents, the errors in excitation energies do not exhibit a linear relationship with the vector potential, making them challenging to correct through simple schemes.

\section{Summary}

rt-TDDFT based on NAOs  is a powerful tool for studying ultrafast dynamics.
However, the conventional implementation of
velocity gauge neglects position-dependent phase within atomic orbitals, leading to significant errors. To address this issue, we propose a hybrid gauge approach preserves essential phase information within NAOs. Moreover, since grid-based integrations for nonlocal potentials are not required, the hybrid gauge is computationally more efficient than the velocity gauge.
We also introduce a simple numerical correction scheme for the velocity gauge in current calculations, which consequently improves the derived spectra and dielectric functions. Nevertheless, this correction cannot fully eliminate errors, especially at strong field intensities.
Overall, the hybrid gauge enables accurate and efficient rt-TDDFT simulations of periodic systems using even small
NAO basis sets. By capturing both linear and nonlinear responses more reliably, it paves the way for studying more complex ultrafast phenomena—such as strong-field excitations, high-harmonic generation, and light-induced phase transitions.

This work was supported by the National Natural Science Foundation of China (Grant Number 12134012),
 the Strategic Priority Research Program of Chinese Academy of Sciences (Grant Number XDB0500201)
and the Innovation Program for Quan-tum Science and Technology Grant Number 2021ZD0301200.
The numerical calculations were performed on the USTC HPC facilities.


\begin{mcitethebibliography}{34}
\providecommand*\natexlab[1]{#1}
\providecommand*\mciteSetBstSublistMode[1]{}
\providecommand*\mciteSetBstMaxWidthForm[2]{}
\providecommand*\mciteBstWouldAddEndPuncttrue
  {\def\EndOfBibitem{\unskip.}}
\providecommand*\mciteBstWouldAddEndPunctfalse
  {\let\EndOfBibitem\relax}
\providecommand*\mciteSetBstMidEndSepPunct[3]{}
\providecommand*\mciteSetBstSublistLabelBeginEnd[3]{}
\providecommand*\EndOfBibitem{}
\mciteSetBstSublistMode{f}
\mciteSetBstMaxWidthForm{subitem}{(\alph{mcitesubitemcount})}
\mciteSetBstSublistLabelBeginEnd
  {\mcitemaxwidthsubitemform\space}
  {\relax}
  {\relax}

\bibitem[Zewail(2000)]{Zewail2000}
Zewail,~A.~H. Femtochemistry: Atomic-scale dynamics of the chemical bond.
  \emph{J. Phys. Chem. A} \textbf{2000}, \emph{104}, 5660--5694\relax
\mciteBstWouldAddEndPuncttrue
\mciteSetBstMidEndSepPunct{\mcitedefaultmidpunct}
{\mcitedefaultendpunct}{\mcitedefaultseppunct}\relax
\EndOfBibitem
\bibitem[Strickland and Mourou(1985)Strickland, and Mourou]{Strickland1985}
Strickland,~D.; Mourou,~G. Compression of amplified chirped optical pulses.
  \emph{Opt. Commun.} \textbf{1985}, \emph{55}, 447--449\relax
\mciteBstWouldAddEndPuncttrue
\mciteSetBstMidEndSepPunct{\mcitedefaultmidpunct}
{\mcitedefaultendpunct}{\mcitedefaultseppunct}\relax
\EndOfBibitem
\bibitem[Krausz and Ivanov(2009)Krausz, and Ivanov]{Krausz2009}
Krausz,~F.; Ivanov,~M. Attosecond physics. \emph{Rev. Mod. Phys.}
  \textbf{2009}, \emph{81}, 163--234\relax
\mciteBstWouldAddEndPuncttrue
\mciteSetBstMidEndSepPunct{\mcitedefaultmidpunct}
{\mcitedefaultendpunct}{\mcitedefaultseppunct}\relax
\EndOfBibitem
\bibitem[Corkum and Krausz(2007)Corkum, and Krausz]{Corkum2007}
Corkum,~P.~B.; Krausz,~F. Attosecond science. \emph{Nat. Phys.} \textbf{2007},
  \emph{3}, 381--387\relax
\mciteBstWouldAddEndPuncttrue
\mciteSetBstMidEndSepPunct{\mcitedefaultmidpunct}
{\mcitedefaultendpunct}{\mcitedefaultseppunct}\relax
\EndOfBibitem
\bibitem[Schultze \latin{et~al.}(2010)Schultze, Fie{\ss}, Karpowicz, Gagnon,
  Korbman, Hofstetter, Neppl, Cavalieri, Komninos, Mercouris, \latin{et~al.}
  others]{Schultze2010}
Schultze,~M.; Fie{\ss},~M.; Karpowicz,~N.; Gagnon,~J.; Korbman,~M.;
  Hofstetter,~M.; Neppl,~S.; Cavalieri,~A.~L.; Komninos,~Y.; Mercouris,~T.;
  others Delay in photoemission. \emph{Science} \textbf{2010}, \emph{328},
  1658--1662\relax
\mciteBstWouldAddEndPuncttrue
\mciteSetBstMidEndSepPunct{\mcitedefaultmidpunct}
{\mcitedefaultendpunct}{\mcitedefaultseppunct}\relax
\EndOfBibitem
\bibitem[L’Huillier and Balcou(1993)L’Huillier, and Balcou]{Huillier1993}
L’Huillier,~A.; Balcou,~P. High-order harmonic generation in rare gases with
  a 1-ps 1053-nm laser. \emph{Phys. Rev. Lett.} \textbf{1993}, \emph{70},
  774\relax
\mciteBstWouldAddEndPuncttrue
\mciteSetBstMidEndSepPunct{\mcitedefaultmidpunct}
{\mcitedefaultendpunct}{\mcitedefaultseppunct}\relax
\EndOfBibitem
\bibitem[Goulielmakis \latin{et~al.}(2008)Goulielmakis, Schultze, Hofstetter,
  Yakovlev, Gagnon, Uiberacker, Aquila, Gullikson, Attwood, Kienberger,
  \latin{et~al.} others]{Goulielmakis2008}
Goulielmakis,~E.; Schultze,~M.; Hofstetter,~M.; Yakovlev,~V.~S.; Gagnon,~J.;
  Uiberacker,~M.; Aquila,~A.~L.; Gullikson,~E.; Attwood,~D.~T.; Kienberger,~R.;
  others Single-cycle nonlinear optics. \emph{Science} \textbf{2008},
  \emph{320}, 1614--1617\relax
\mciteBstWouldAddEndPuncttrue
\mciteSetBstMidEndSepPunct{\mcitedefaultmidpunct}
{\mcitedefaultendpunct}{\mcitedefaultseppunct}\relax
\EndOfBibitem
\bibitem[Z{\"u}rch \latin{et~al.}(2014)Z{\"u}rch, Rothhardt, H{\"a}drich,
  Demmler, Krebs, Limpert, T{\"u}nnermann, Guggenmos, Kleineberg, and
  Spielmann]{zurch2014}
Z{\"u}rch,~M.; Rothhardt,~J.; H{\"a}drich,~S.; Demmler,~S.; Krebs,~M.;
  Limpert,~J.; T{\"u}nnermann,~A.; Guggenmos,~A.; Kleineberg,~U.; Spielmann,~C.
  Real-time and sub-wavelength ultrafast coherent diffraction imaging in the
  extreme ultraviolet. \emph{Sci. Rep.} \textbf{2014}, \emph{4}, 7356\relax
\mciteBstWouldAddEndPuncttrue
\mciteSetBstMidEndSepPunct{\mcitedefaultmidpunct}
{\mcitedefaultendpunct}{\mcitedefaultseppunct}\relax
\EndOfBibitem
\bibitem[Chini \latin{et~al.}(2014)Chini, Zhao, and Chang]{Chini2014}
Chini,~M.; Zhao,~K.; Chang,~Z. The generation, characterization and
  applications of broadband isolated attosecond pulses. \emph{Nat. Photonics}
  \textbf{2014}, \emph{8}, 178--186\relax
\mciteBstWouldAddEndPuncttrue
\mciteSetBstMidEndSepPunct{\mcitedefaultmidpunct}
{\mcitedefaultendpunct}{\mcitedefaultseppunct}\relax
\EndOfBibitem
\bibitem[Runge and Gross(1984)Runge, and Gross]{Runge1984}
Runge,~E.; Gross,~E. K.~U. Density-Functional Theory for Time-Dependent
  Systems. \emph{Phys. Rev. Lett.} \textbf{1984}, \emph{52}, 997--1000\relax
\mciteBstWouldAddEndPuncttrue
\mciteSetBstMidEndSepPunct{\mcitedefaultmidpunct}
{\mcitedefaultendpunct}{\mcitedefaultseppunct}\relax
\EndOfBibitem
\bibitem[Yabana and Bertsch(1996)Yabana, and Bertsch]{Yabana1996}
Yabana,~K.; Bertsch,~G. Time-dependent local-density approximation in real
  time. \emph{Phys. Rev. B} \textbf{1996}, \emph{54}, 4484\relax
\mciteBstWouldAddEndPuncttrue
\mciteSetBstMidEndSepPunct{\mcitedefaultmidpunct}
{\mcitedefaultendpunct}{\mcitedefaultseppunct}\relax
\EndOfBibitem
\bibitem[Bertsch \latin{et~al.}(2000)Bertsch, Iwata, Rubio, and
  Yabana]{Bertsch2000}
Bertsch,~G.~F.; Iwata,~J.-I.; Rubio,~A.; Yabana,~K. Real-space, real-time
  method for the dielectric function. \emph{Phys. Rev. B} \textbf{2000},
  \emph{62}, 7998\relax
\mciteBstWouldAddEndPuncttrue
\mciteSetBstMidEndSepPunct{\mcitedefaultmidpunct}
{\mcitedefaultendpunct}{\mcitedefaultseppunct}\relax
\EndOfBibitem
\bibitem[Marques and Gross(2004)Marques, and Gross]{Marques2004}
Marques,~M.~A.; Gross,~E.~K. Time-dependent density functional theory.
  \emph{Annu. Rev. Phys. Chem.} \textbf{2004}, \emph{55}, 427--455\relax
\mciteBstWouldAddEndPuncttrue
\mciteSetBstMidEndSepPunct{\mcitedefaultmidpunct}
{\mcitedefaultendpunct}{\mcitedefaultseppunct}\relax
\EndOfBibitem
\bibitem[Yabana and Bertsch(1999)Yabana, and Bertsch]{Yabana1999}
Yabana,~K.; Bertsch,~G. Time-dependent local-density approximation in real
  time: application to conjugated molecules. \emph{Int. J. Quantum Chem.}
  \textbf{1999}, \emph{75}, 55--66\relax
\mciteBstWouldAddEndPuncttrue
\mciteSetBstMidEndSepPunct{\mcitedefaultmidpunct}
{\mcitedefaultendpunct}{\mcitedefaultseppunct}\relax
\EndOfBibitem
\bibitem[Castro \latin{et~al.}(2004)Castro, Marques, Alonso, Bertsch, and
  Rubio]{Castro2004td}
Castro,~A.; Marques,~M.~A.; Alonso,~J.~A.; Bertsch,~G.~F.; Rubio,~A. Excited
  states dynamics in time-dependent density functional theory: high-field
  molecular dissociation and harmonic generation. \emph{Eur. Phys. J. D}
  \textbf{2004}, \emph{28}, 211--218\relax
\mciteBstWouldAddEndPuncttrue
\mciteSetBstMidEndSepPunct{\mcitedefaultmidpunct}
{\mcitedefaultendpunct}{\mcitedefaultseppunct}\relax
\EndOfBibitem
\bibitem[Otobe \latin{et~al.}(2008)Otobe, Yamagiwa, Iwata, Yabana, Nakatsukasa,
  and Bertsch]{Otobe2008}
Otobe,~T.; Yamagiwa,~M.; Iwata,~J.-I.; Yabana,~K.; Nakatsukasa,~T.; Bertsch,~G.
  First-principles electron dynamics simulation for optical breakdown of
  dielectrics under an intense laser field. \emph{Phys. Rev. B} \textbf{2008},
  \emph{77}, 165104\relax
\mciteBstWouldAddEndPuncttrue
\mciteSetBstMidEndSepPunct{\mcitedefaultmidpunct}
{\mcitedefaultendpunct}{\mcitedefaultseppunct}\relax
\EndOfBibitem
\bibitem[Provorse and Isborn(2016)Provorse, and Isborn]{Provorse2016}
Provorse,~M.~R.; Isborn,~C.~M. Electron dynamics with real-time time-dependent
  density functional theory. \emph{Int. J. Quantum Chem.} \textbf{2016},
  \emph{116}, 739--749\relax
\mciteBstWouldAddEndPuncttrue
\mciteSetBstMidEndSepPunct{\mcitedefaultmidpunct}
{\mcitedefaultendpunct}{\mcitedefaultseppunct}\relax
\EndOfBibitem
\bibitem[Sato \latin{et~al.}(2014)Sato, Yabana, Shinohara, Otobe, and
  Bertsch]{Sato2014}
Sato,~S.; Yabana,~K.; Shinohara,~Y.; Otobe,~T.; Bertsch,~G.~F. Numerical
  pump-probe experiments of laser-excited silicon in nonequilibrium phase.
  \emph{Phys. Rev. B} \textbf{2014}, \emph{89}, 064304\relax
\mciteBstWouldAddEndPuncttrue
\mciteSetBstMidEndSepPunct{\mcitedefaultmidpunct}
{\mcitedefaultendpunct}{\mcitedefaultseppunct}\relax
\EndOfBibitem
\bibitem[Yabana \latin{et~al.}(2012)Yabana, Sugiyama, Shinohara, Otobe, and
  Bertsch]{Yabana2012}
Yabana,~K.; Sugiyama,~T.; Shinohara,~Y.; Otobe,~T.; Bertsch,~G. Time-dependent
  density functional theory for strong electromagnetic fields in crystalline
  solids. \emph{Phys. Rev. B} \textbf{2012}, \emph{85}, 045134\relax
\mciteBstWouldAddEndPuncttrue
\mciteSetBstMidEndSepPunct{\mcitedefaultmidpunct}
{\mcitedefaultendpunct}{\mcitedefaultseppunct}\relax
\EndOfBibitem
\bibitem[Lin \latin{et~al.}(2024)Lin, Ren, Liu, and He]{NAO-review}
Lin,~P.; Ren,~X.; Liu,~X.; He,~L. Ab initio electronic structure calculations
  based on numerical atomic orbitals: Basic fomalisms and recent progresses.
  \emph{WIREs Comput. Mol. Sci.} \textbf{2024}, \emph{14}, e1687\relax
\mciteBstWouldAddEndPuncttrue
\mciteSetBstMidEndSepPunct{\mcitedefaultmidpunct}
{\mcitedefaultendpunct}{\mcitedefaultseppunct}\relax
\EndOfBibitem
\bibitem[Pemmaraju \latin{et~al.}(2018)Pemmaraju, Vila, Kas, Sato, Rehr,
  Yabana, and Prendergast]{PEMMARAJU201830}
Pemmaraju,~C.~D.; Vila,~F.~D.; Kas,~J.~J.; Sato,~S.~A.; Rehr,~J.~J.;
  Yabana,~K.; Prendergast,~D. Velocity-gauge real-time TDDFT within a numerical
  atomic orbital basis set. \emph{Computer Physics Communications}
  \textbf{2018}, \emph{226}, 30--38\relax
\mciteBstWouldAddEndPuncttrue
\mciteSetBstMidEndSepPunct{\mcitedefaultmidpunct}
{\mcitedefaultendpunct}{\mcitedefaultseppunct}\relax
\EndOfBibitem
\bibitem[Lian \latin{et~al.}(2018)Lian, Hu, Guan, and Meng]{TDAP}
Lian,~C.; Hu,~S.~Q.; Guan,~M.~X.; Meng,~S. Momentum-resolved {TDDFT} algorithm
  in atomic basis for real time tracking of electronic excitation. \emph{J.
  Chem. Phys.} \textbf{2018}, \emph{149}, 154104\relax
\mciteBstWouldAddEndPuncttrue
\mciteSetBstMidEndSepPunct{\mcitedefaultmidpunct}
{\mcitedefaultendpunct}{\mcitedefaultseppunct}\relax
\EndOfBibitem
\bibitem[Hekele \latin{et~al.}(2021)Hekele, Yao, Kanai, Blum, and
  Kratzer]{Aims-tddft}
Hekele,~J.; Yao,~Y.; Kanai,~Y.; Blum,~V.; Kratzer,~P. All-electron real-time
  and imaginary-time time-dependent density functional theory within a numeric
  atom-centered basis function framework. \emph{J. Chem. Phys.} \textbf{2021},
  \emph{155}, 154801\relax
\mciteBstWouldAddEndPuncttrue
\mciteSetBstMidEndSepPunct{\mcitedefaultmidpunct}
{\mcitedefaultendpunct}{\mcitedefaultseppunct}\relax
\EndOfBibitem
\bibitem[He \latin{et~al.}(2024)He, Chen, Ren, Meng, and He]{Hefx2024}
He,~F.; Chen,~D.; Ren,~X.; Meng,~S.; He,~L. Ultrafast shift current dynamics in
  {WS}$_2$ monolayer. \emph{Phys. Rev. Res.} \textbf{2024}, \emph{6},
  013123\relax
\mciteBstWouldAddEndPuncttrue
\mciteSetBstMidEndSepPunct{\mcitedefaultmidpunct}
{\mcitedefaultendpunct}{\mcitedefaultseppunct}\relax
\EndOfBibitem
\bibitem[Lian \latin{et~al.}(2018)Lian, Guan, Hu, Zhang, and Meng]{Lian2018}
Lian,~C.; Guan,~M.; Hu,~S.; Zhang,~J.; Meng,~S. Photoexcitation in Solids:
  First-Principles Quantum Simulations by Real-Time TDDFT. \emph{Advanced
  Theory and Simulations} \textbf{2018}, \emph{1}, 1800055\relax
\mciteBstWouldAddEndPuncttrue
\mciteSetBstMidEndSepPunct{\mcitedefaultmidpunct}
{\mcitedefaultendpunct}{\mcitedefaultseppunct}\relax
\EndOfBibitem
\bibitem[You \latin{et~al.}(2021)You, Chen, Lian, Zhang, and Meng]{You2021}
You,~P.; Chen,~D.; Lian,~C.; Zhang,~C.; Meng,~S. First-principles dynamics of
  photoexcited molecules and materials towards a quantum description.
  \emph{WIREs Comput. Mol. Sci.} \textbf{2021}, \emph{11}, e1492\relax
\mciteBstWouldAddEndPuncttrue
\mciteSetBstMidEndSepPunct{\mcitedefaultmidpunct}
{\mcitedefaultendpunct}{\mcitedefaultseppunct}\relax
\EndOfBibitem
\bibitem[Kleinman and Bylander(1982)Kleinman, and Bylander]{Kleinman1982}
Kleinman,~L.; Bylander,~D.~M. Efficacious Form for Model Pseudopotentials.
  \emph{Phys. Rev. Lett.} \textbf{1982}, \emph{48}, 1425--1428\relax
\mciteBstWouldAddEndPuncttrue
\mciteSetBstMidEndSepPunct{\mcitedefaultmidpunct}
{\mcitedefaultendpunct}{\mcitedefaultseppunct}\relax
\EndOfBibitem
\bibitem[Soler \latin{et~al.}(2002)Soler, Artacho, Gale, Garc{\'i}a, Junquera,
  Ordej{\'o}n, and S{\'a}nchez-Portal]{Soler/etal:2002}
Soler,~J.~M.; Artacho,~E.; Gale,~J.~D.; Garc{\'i}a,~A.; Junquera,~J.;
  Ordej{\'o}n,~P.; S{\'a}nchez-Portal,~D. The SIESTA method for ab initio
  order-N materials simulation. \emph{J. Phys.: Condens. Matter} \textbf{2002},
  \emph{14}, 2745\relax
\mciteBstWouldAddEndPuncttrue
\mciteSetBstMidEndSepPunct{\mcitedefaultmidpunct}
{\mcitedefaultendpunct}{\mcitedefaultseppunct}\relax
\EndOfBibitem
\bibitem[Li \latin{et~al.}(2016)Li, Liu, Chen, Lin, Ren, Lin, Yang, and
  He]{ABACUS2}
Li,~P.; Liu,~X.; Chen,~M.; Lin,~P.; Ren,~X.; Lin,~L.; Yang,~C.; He,~L.
  Large-scale ab initio simulations based on systematically improvable atomic
  basis. \emph{Comput. Mater. Sci.} \textbf{2016}, \emph{112}, 503--517\relax
\mciteBstWouldAddEndPuncttrue
\mciteSetBstMidEndSepPunct{\mcitedefaultmidpunct}
{\mcitedefaultendpunct}{\mcitedefaultseppunct}\relax
\EndOfBibitem
\bibitem[Lin \latin{et~al.}(2021)Lin, Ren, and He]{Lin_orbital}
Lin,~P.; Ren,~X.; He,~L. Strategy for constructing compact numerical atomic
  orbital basis sets by incorporating the gradients of reference wavefunctions.
  \emph{Phys. Rev. B} \textbf{2021}, \emph{103}, 235131\relax
\mciteBstWouldAddEndPuncttrue
\mciteSetBstMidEndSepPunct{\mcitedefaultmidpunct}
{\mcitedefaultendpunct}{\mcitedefaultseppunct}\relax
\EndOfBibitem
\bibitem[Perdew \latin{et~al.}(1996)Perdew, Burke, and Ernzerhof]{GGA}
Perdew,~J.~P.; Burke,~K.; Ernzerhof,~M. Generalized Gradient Approximation Made
  Simple. \emph{Phys. Rev. Lett.} \textbf{1996}, \emph{77}, 3865--3868\relax
\mciteBstWouldAddEndPuncttrue
\mciteSetBstMidEndSepPunct{\mcitedefaultmidpunct}
{\mcitedefaultendpunct}{\mcitedefaultseppunct}\relax
\EndOfBibitem
\bibitem[Scherpelz \latin{et~al.}(2016)Scherpelz, Govoni, Hamada, and
  Galli]{SG15}
Scherpelz,~P.; Govoni,~M.; Hamada,~I.; Galli,~G. Implementation and Validation
  of Fully Relativistic GW Calculations: Spin–Orbit Coupling in Molecules,
  Nanocrystals, and Solids. \emph{J. Chem. Theory Comput.} \textbf{2016},
  \emph{12}, 3523--3544\relax
\mciteBstWouldAddEndPuncttrue
\mciteSetBstMidEndSepPunct{\mcitedefaultmidpunct}
{\mcitedefaultendpunct}{\mcitedefaultseppunct}\relax
\EndOfBibitem
\bibitem[Sato and Yabana(2014)Sato, and Yabana]{ARTED}
Sato,~S.~A.; Yabana,~K. Maxwell + TDDFT multi-scale simulation for laser-matter
  interactions. \emph{J. Adv. Simul. Sci. Eng.} \textbf{2014}, \emph{1},
  98--110\relax
\mciteBstWouldAddEndPuncttrue
\mciteSetBstMidEndSepPunct{\mcitedefaultmidpunct}
{\mcitedefaultendpunct}{\mcitedefaultseppunct}\relax
\EndOfBibitem
\end{mcitethebibliography}

\providecommand{\latin}[1]{#1}
\makeatletter
\providecommand{\doi}
  {\begingroup\let\do\@makeother\dospecials
  \catcode`\{=1 \catcode`\}=2 \doi@aux}
\providecommand{\doi@aux}[1]{\endgroup\texttt{#1}}
\makeatother
\providecommand*\mcitethebibliography{\thebibliography}
\csname @ifundefined\endcsname{endmcitethebibliography}
  {\let\endmcitethebibliography\endthebibliography}{}

\end{document}